\documentclass[useAMS,usenatbib]{mn2e}
\usepackage{amssymb,amsmath,epsfig,times, natbib,color}
\pdfoutput=1

\title[A2199 outskirts]{A high coverage view of the thermodynamics and metal abundance in the outskirts of the nearby galaxy cluster Abell 2199 }
\author[M. S. Mirakhor et al] {\parbox[]{6.5in}{
      M. S. Mirakhor$^1$\thanks{Email: 
    msm0033@uah.edu}, {S. A. Walker$^1$}}\\
     \footnotesize
     $^1$Department of Physics and Astronomy, The University of Alabama in Huntsville, 301 Sparkman Drive NW, Huntsville, AL 35899, USA \\
}

\date{}

\begin{document}

\maketitle

\begin{abstract}
We present a joint \textit{Suzaku} and \textit{XMM-Newton} analysis of the outskirts of the nearby galaxy cluster Abell 2199, the only nearby galaxy cluster to be observed with near complete azimuthal coverage with \textit{Suzaku}. Using the \textit{XMM-Newton} observations to correct for the effects of gas clumping, we find that the azimuthally averaged entropy profile in the outskirts follows a power law with a slope of $1.20 \pm 0.23$, statistically consistent with a slope of 1.1 predicted by non-radiative simulations for purely gravitational hierarchical structure formation. However, when divided into 10 sectors, the entropy shows significant azimuthal variation, with some sectors lying below the baseline level. The azimuthally averaged gas mass fraction is found to agree with the cosmic mean baryon fraction. The metal abundance in the outskirts is found to be consistent with being uniform in all directions and it has an average value of $0.29_{-0.03}^{+0.03}\,Z_{\odot}$, consistent with the gas accreting onto clusters being pre-enriched with metals.
\end{abstract}

\section{Introduction}

The low and stable particle background of the \textit{Suzaku} X-ray observatory allowed the first temperature measurements to be made of the intracluster medium (ICM) near the virial radius\footnote{In this paper we follow the common practice of refering to $r_{200}$ (the radius within which the mean density is 200 times the critical density of the universe) as the virial radius.} (see \citealt{Walker2019} for a review). However a complete, detailed view of the outskirts of the ICM was challenging owing to the large point spread function (PSF) of \textit{Suzaku} (a half power diamater of around 2 arcmin) and its modest field of view. \textit{Suzaku} was able to achieve high azimuthal coverage of higher redshift clusters ($z>0.07$) with mosaics of around 4 observations, however the large PSF meant that these clusters could only be resolved into a small number of annuli, and only around 4 sectors (e.g. A1689, \citealt{Kawaharada2010}; A2029, \citealt{Walker2012_A2029}; PKS0745-171, \citealt{Walker2012_PKS0745}b). 

To get around \textit{Suzaku}'s large PSF, nearby clusters were also observed in strips from the centre to the outskirts, such as the Perseus cluster \citep{Simionescu2011,Urban2014}, the Centaurus cluster \citep{Walker2013_Centaurus}, the Coma cluster \citep{simionescu2013thermodynamics}, and the Virgo cluster \citep{Simionescu17}. However, the azimuthal coverage of these mosaics was low due to the infeasibly large amount of observing time needed to cover the clusters in their entirety. 

Here we present the \textit{Suzaku} mosaic observations of the nearby galaxy cluster Abell 2199, the only nearby galaxy cluster to be observed with near complete azimuthal coverage by \textit{Suzaku}. A2199 is a regular, bright, and rich galaxy cluster at a redshift of 0.03. A2199 with Abell 2197 and several nearby groups forms the A2199 supercluster \citep{rines2001x,lee2015galaxy}. All infalling groups are locating far beyond the virial radius of A2199 \citep{rines2001x}. X-ray studies \citep[e.g.][]{rines2001x,kawaharada2010a2199} found that the inner region of A2199 has not undergone recent major merger events.  

Abell 2199's large spatial extent means that it can be divided into many more spatially resolved sectors than the higher redshift clusters \textit{Suzaku} has studied. This is important because simulations predict that the outskirts of galaxy clusters should become increasingly asymmetric in their outskirts (e.g. \citealt{Roncarelli2006}, \citealt{vazza2011scatter}, \citealt{avestruz2014testing}, \citealt{lau2015mass}). Congruent archival \textit{XMM-Newton} observations for the majority of the \textit{Suzaku} observations are available, and allow for an independent measure of the gas density and, importantly, allow the clumping factor of the intracluster medium to be measured. \textit{XMM-Newton}'s higher spatial resolution allows gas clumping to be measured through the ratio of the mean to median surface brightness, as has been demonstrated by e.g. \citet{eckert2015gas}, \cite{Tchernin2016}, \citet{ghirardini2018xmm}. 

Using \textit{Suzaku}, we can also measure the spatial variation of the metal abundance of the ICM in the outskirts. Measurements of the metal abundance in the outskirts of galaxy clusters are a powerful probe of feedback physics (\cite{Werner2013, urban2017uniform, Simionescu17, Biffi2018, Mernier2018}). This is because the metal distribution in the ICM is strongly dependent on the history of chemical enrichment. Cosmological hydrodynamical simulations \citep{biffi2017history,Biffi2018} found that AGN feedback at early times ($z > 2$) is effective at removing pre-enriched gas from galaxies, which leads to metals being spread uniformly throughout the cluster outskirts. On the other hand, late-time enrichment leads to inhomogeneous distributions of ICM metals that rapidly decline with cluster radius.

Throughout this paper, we adopt a $\Lambda$ CDM cosmology with $\Omega_{\rm{m}}=0.3$, $\Omega_{\rm{\Lambda}}=0.7$, and $H_0=100\,h_{100}$ km s$^{-1}$ Mpc$^{-1}$ with $h_{100}=0.7$. At the redshift of Abell 2199 ($z=0.03$), 1 arcmin corresponds to 36 kpc.

\section{Data}
\label{sec: data}
In this work, a joint analysis of \textit{Suzaku} and \textit{XMM-Newton} observations were used to analyse the thermodynamic properties of A2199 up to the virial radius. The \textit{Suzaku} mosaic consists of 22 pointings, whereas the \textit{XMM-Newton} mosaic consists of 12 pointings. The details regarding the \textit{Suzaku} and \textit{XMM-Newton} observations are summarized in Table \ref{tab: Suzaku_observations} and Table \ref{tab: xmm_observations}, respectively.

\subsection{XMM Data Reduction}
\label{section:XMMreduction}
The XMM data were reduced using the \textit{XMM-Newton} Science Analysis System, XMM-SAS v18.0, and the calibration files, following the procedures illustrated in the Extended Source Analysis Software manual \citep[ESAS,][]{snowden2014cookbook}. The data were examined for anomalous CCDs, and any affected CCDs were removed from the analysis. The ESAS tool \textit{cheese} and also the \textit{Chandra} tool \textit{wavdetect} were used to mask point sources and extended  substructures that contaminated the field of view. For all three detectors, the spectra and response files were created by running the \textit{mos-spectra} and \textit{pn-spectra} tasks. These files were then used to create the quiescent particle background spectra and images using the \textit{mos-back} and \textit{pn-back}. The data were also examined for residual soft proton contamination using the ESAS task \textit{proton}.

\begin{figure*}
	\includegraphics[width=\textwidth]{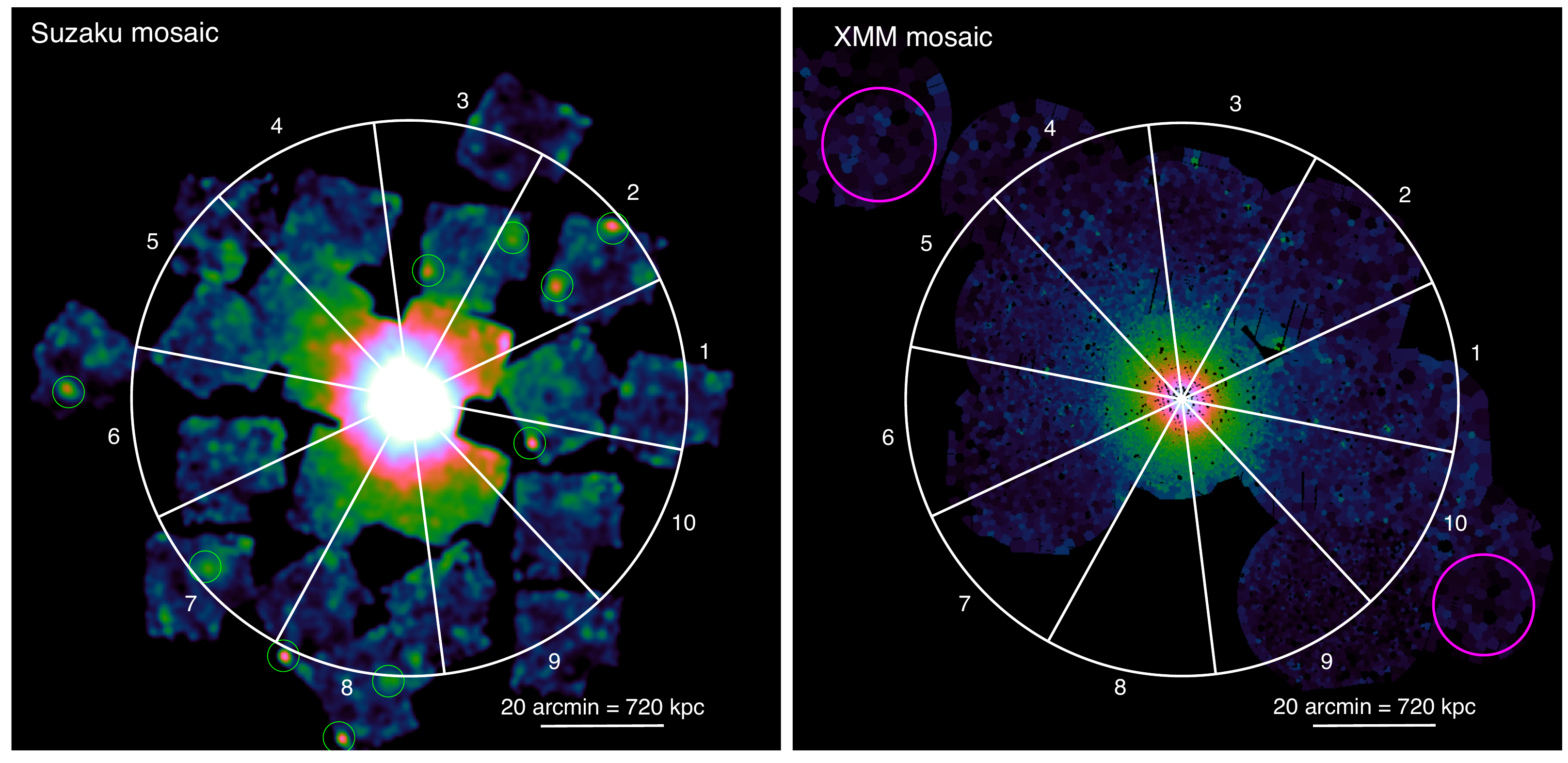}
	\caption{\textit{Left}: \textit{Suzaku} mosaic of A2199 with sectors overplotted. Green circles show the brightest point sources which were excluded from the analysis. \textit{Right}: Voronoi tesselated, background subtracted and exposure corrected \textit{XMM-Newton} mosaic of A2199. The magenta circles show the regions in which the local background counts were extracted.}
	\label{fig: A2199_mosaics}
\end{figure*}

The analysis procedures described above created all of the required components for a background-subtracted and exposure-corrected image. After weighting each detector by its relative effective area, these components from all three XMM detectors and 12 pointings were then combined and adaptively smoothed into a single image. Furthermore, we applied a Voronoi tessellation algorithm using the method of \citet{diehl2006adaptive} to create an adaptively binned image, where each bin contains at least 20 counts. The resulting background-subtracted and exposure-corrected image of A2199 is shown in the right panel of Fig. \ref{fig: A2199_mosaics}. The X-ray image of A2199 was created in the 0.7-1.2 energy band in order to maximize the signal-to-noise ratio in the outskirts \citep[e.g.][]{Tchernin2016}. 

\begin{figure}
	\includegraphics[width=\columnwidth]{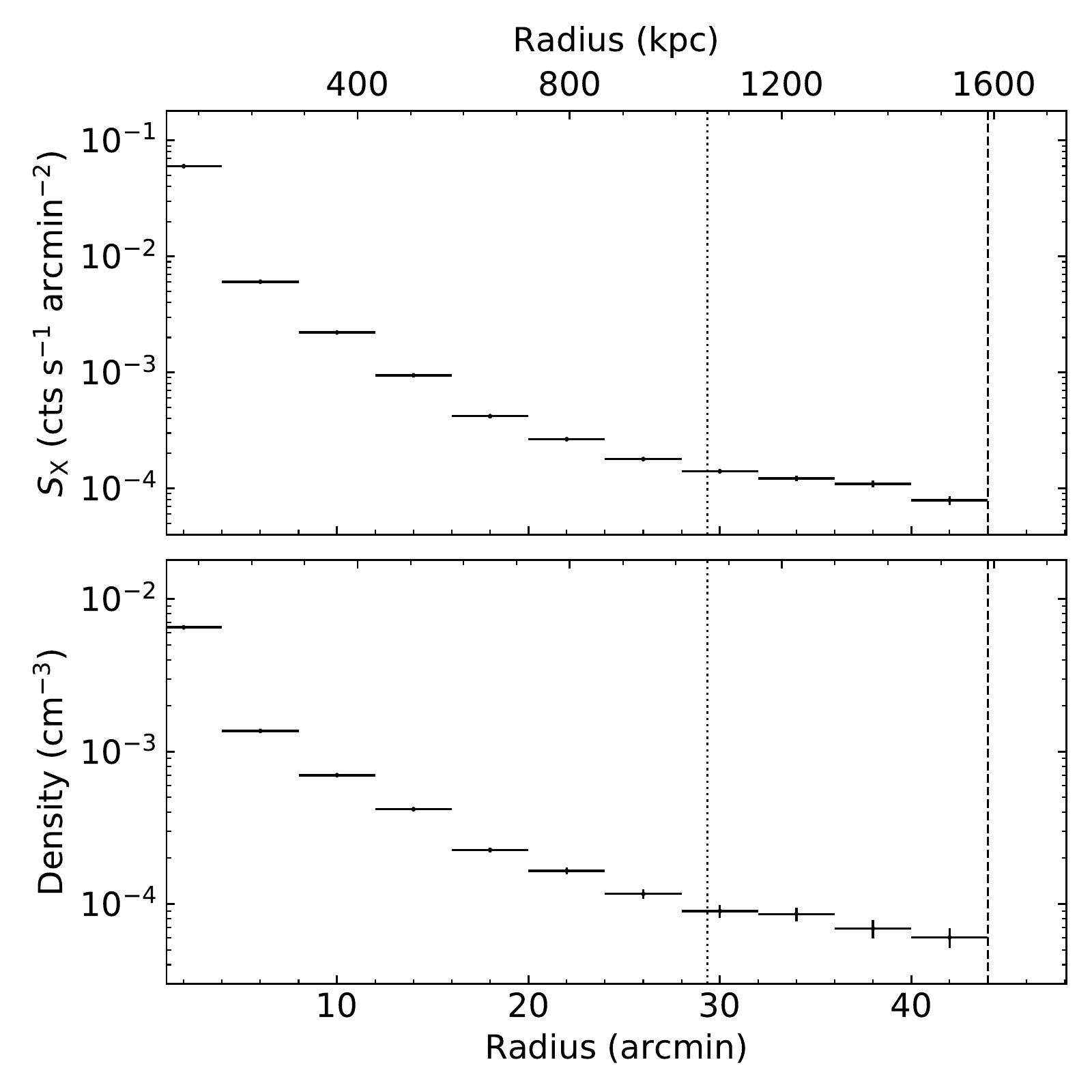}
	\caption{\textit{Top}: Azimuthal median of the background-subtracted surface brightness profile for the A2199 cluster in the 0.7$-$1.2 keV energy band. \textit{Bottom}: Density profile obtained from the deprojection of the surface brightness radial profile using the onion peeling technique. The error bars are the 1$\sigma$ percentiles computed using a Monte Carlo technique. The vertical dotted and dashed lines represent the $r_{500}$ and $r_{200}$ radii, respectively.}
	\label{fig: A2199_sb_ne_azav}
\end{figure}

The surface brightness profile for the mosaicked image of A2199 was then extracted in concentric annuli centered at the cluster centre (RA, Dec.) = (16:28:38.21, +39:33:02.31). This corresponds to the location of the peak X-ray flux in the cluster. The radial profile of the surface brightness is shown in the upper panel of Fig. \ref{fig: A2199_sb_ne_azav}. In the lower panel of this figure, we present the radial profile of the azimuthally averaged electron density of A2199. The profile is recovered from the deprojection of the median surface brightness profile using the onion peeling technique \citep{ettori2010mass}, and assuming that the ICM plasma is spherical symmetry. To convert from surface brightness to density, we use the following widely used approach \citep{Eckert2012,Tchernin2016,ghirardini2018xmm,Ghirardini2019,Walker2020} Using XSPEC and the response files for \textit{XMM-Newton}, the conversion factor between APEC normalization and X-ray count rate in the 0.7$-$1.2 keV band is found. In the 0.7$-$1.2 keV band, this is largely independent of the temperature of the gas, so this allows a direct conversion from the deprojected X-ray surface brightness profile to a deprojected profile of APEC normalization. The APEC normalization is related to the gas density by the equation:

\begin{equation}
 {\rm Norm} = \frac{10^{-14}}{4 \pi [d_A(1+z)]^2 }\int n_e n_H dV
 \label{equ: APEC_norm}
\end{equation}
where the electron and ion number densities are $n_e$ and $n_H$ respectively (for a fully ionized plasma these are related by $n_e=1.17n_H$, \citealt{Asplund2009}), the angular diameter distance to the cluster is $d_A$, and $z$ is the cluster redshift. The deprojected APEC normalizations are then converted to deprojected density in the usual fashion, assuming spherical symmetry and a constant density in each shell, by calculating the projected volumes, V, of each shell in the 2D annuli. When performing this conversion, we used a column density of 0.08$\times$ $10^{22}$ cm$^{-2}$ (from \citealt{LABsurvey}), and using the abundance tables of \citet{Asplund2009}.

\begin{figure}
	\includegraphics[width=\columnwidth]{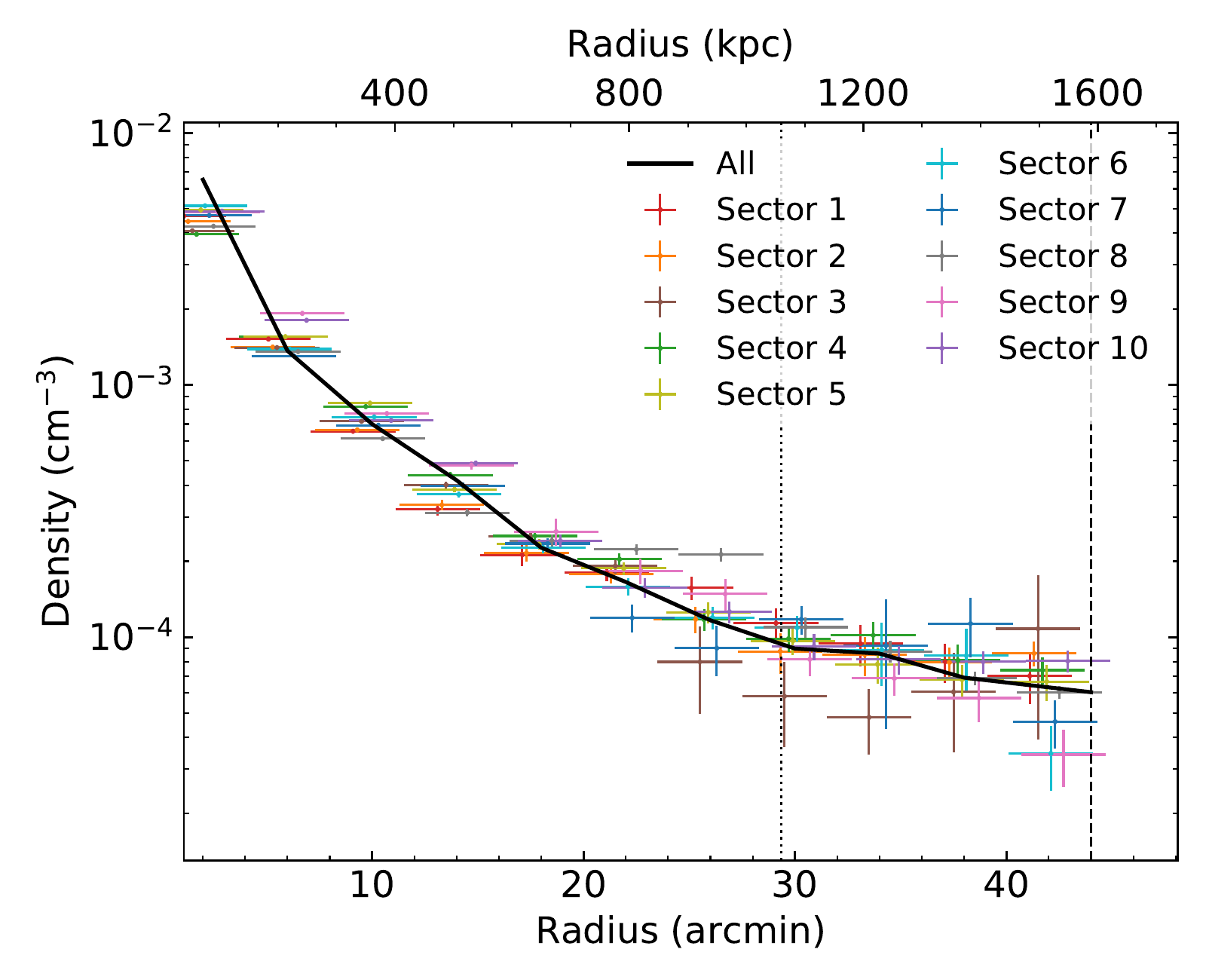}
	\caption{Density profiles in the 10 angular sectors recovered from the deprojection of the X-ray surface brightness using the onion peeling technique. The error bars are the 1$\sigma$ percentiles computed using a Monte Carlo technique. The thick black line shows the azimuthally averaged density profile. The vertical dotted and dashed lines mark the $r_{500}$ and $r_{200}$ radii, respectively. We introduce small offsets between data points at the same radius to aid readability.}
	\label{fig: A2199_density}
\end{figure}

The density profile is also derived in each of the 10 azimuthal sectors shown in Fig. \ref{fig: A2199_mosaics}, taking advantage of the high signal-to-noise ratio of our data. In this azimuthal sector analysis, any small gaps found in the \textit{XMM-Newton} mosaic are linearly interpolated over. For sector 8, where there is no \textit{XMM-Newton} coverage, we used the \textit{Suzaku} data to derive the gas density. Fig. \ref{fig: A2199_density} shows the density profiles in the azimuthal sectors of the A2199 cluster. 

\begin{figure}
	\includegraphics[width=\columnwidth]{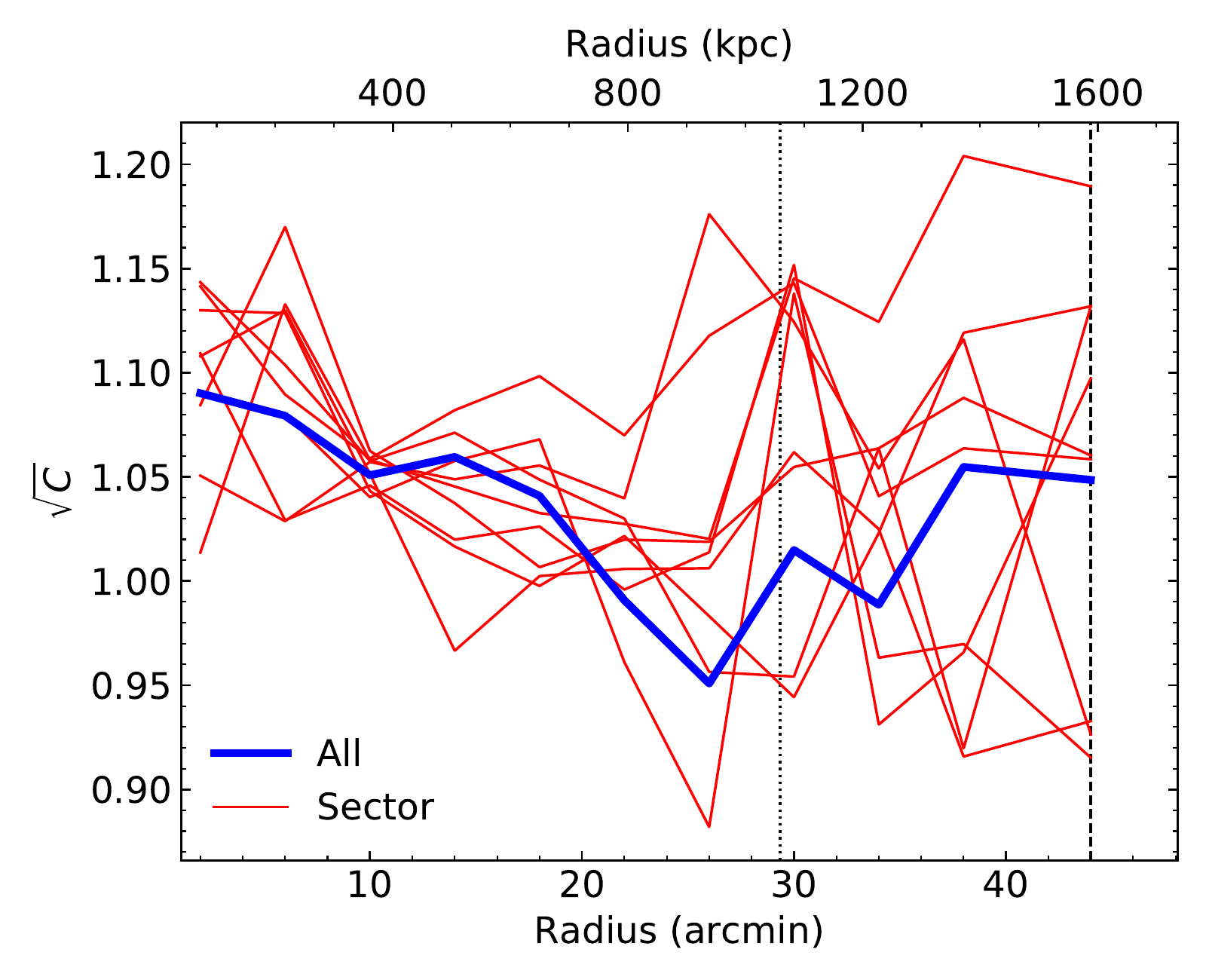}
	\caption{Profiles of the clumping factor for the azimuthally averaged profile (\textit{blue}) and azimuthal sectors (\textit{red}). The vertical dotted and dashed lines mark, respectively, the $r_{500}$ and $r_{200}$ radii.}
	\label{fig: clumping}
\end{figure}

The derived gas density profiles, however, could be biased by presence of gas clumping, which become increasingly important near the virial radius \citep{Walker2019}. We estimated the level of this possible bias directly from the mosaicked images by dividing the mean surface brightness to the median surface brightness in each annulus region, following \citet{eckert2015gas}. The inferred clumping factor, $\sqrt{C}$, is plotted in Fig \ref{fig: clumping}. The estimated values of  the  clumping  factor are less than 1.1 over the entire cluster-radial range, implying that the clumping bias is very mild in A2199.

Typically, the radial profile of the X-ray surface brightness is subject to various sources of systematic uncertainties. The choice of the local background region is the major source of systematic uncertainty in the imaging-data analysis. To address this issue, we chose two local background regions in two different positions outside the virial radius of A2199, as shown in the magenta regions in Fig. \ref{fig: A2199_mosaics}. Furthermore, we take into account uncertainty due to background fluctuations by considering a $\pm 5\%$ uncertainty of the background level as an additional uncertainty in the radial surface brightness profile. Measurements presented in this work take account of the systematic uncertainties.

\subsection{\textit{Suzaku} Data Reduction}
\label{section:Suzaku}

\textit{Suzaku} observed Abell 2199 with 22 observations covering the whole cluster with high azimuthal coverage in the outskirts, and these observations are tabulated in table \ref{tab: Suzaku_observations}, and a background subtracted, exposure corrected mosaic of the observations is shown in the left hand panel of Fig. \ref{fig: A2199_mosaics}. The reduction of the \textit{Suzaku} data follows the methods described in \citet{Walker2012_A2029}a, \citet{Walker2012_PKS0745}b and \citet{Walker2013_Centaurus}. Each observation was reprocessed using the tool \textsc{aepipeline}, performing the standard cleaning described in the \textit{Suzaku} ABC guide\footnote{https://heasarc.gsfc.nasa.gov/docs/suzaku/analysis/abc/}. The calibration regions at the edges of the detectors were removed, and the small strip of the XI0 detector that is defective (from a micrometeorite hit) was excluded from the analysis. 

Non X-ray background files (NXBs) were produced using the tool \textsc{xisnxbgen} using the latest calibration files. To a make the mosaiced image, images were extracted in the soft (0.5-2.0 keV) band, and exposure maps were produced using \textsc{xissim}. The NXB image was subtracted from the image from each observation, which was then exposure corrected using the exposure map, before being mosaiced together to produce the image shown in the left panel of Fig. \ref{fig: A2199_mosaics}. We use all three available detectors (XIS0, XIS1 and XIS3).

For spectral fitting, spectra were extracted from each annulus studied. Ancillary response files (ARFs) were produced using the tool \textsc{xissimarfgen}, while RMFs were produced using \textsc{xisrmfgen}.


The light curves of every observation were checked to ensure no flaring occurred during the clean sections of the observations. We checked for solar wind charge exchange (SWCX) contamination by finding the proton flux observed by the WIND spacecraft's solar wind experiment instrument\footnote{https://wind.nasa.gov/}. We found the proton flux to be low and stable during the \textit{Suzaku} observations we use. Previous works \citep{Fujimoto2007, Yoshino2009} have found that SWCX is negligible when the proton flux is below $4\times10^8$ cm$^{-2}$ s$^{-1}$, which we find to be the case.

\subsubsection{\textit{Suzaku} background modelling}
\label{Suzaku_bkg_mod}
We follow the spectral analysis and background modelling procedures described in \citet{Walker2013_Centaurus}. For the background model, we model the cosmic X-ray background (CXB) as a powerlaw with powerlaw index 1.4. The congruent \textit{XMM-Newton} observations allow point sources to be identified uniformly in the majority of the \textit{Suzaku} observations (all of the sectors except sector 8) down to a threshold flux of $1\times10^{-14}$ erg s$^{-1}$ cm$^{-2}$. We then integrate the known cumulative flux distribution of point sources (modelled as a double power law in \citealt{Moretti2003}) down to this threshold flux, which allows us to calculate the expected level of the unresolved cosmic X-ray background below a threshold flux of $1\times10^{-14}$ erg s$^{-1}$ cm$^{-2}$. 

For a given spectral extraction region, we can then calculate the expected variance of the unresolved CXB level following \citet{Walker2013_Centaurus} and include this uncertainty in our systematic error budget. Because Abell 2199 is very nearby ($z=0.03$), we can use large spectral extraction areas, which reduces the variance in the unresolved CXB level.

For the soft background, we use an annular region from the ROSAT All Sky Survey (RASS) reaching from 60-120 arcmins, which we divide into 4 quadrants to assess the annular variation of the soft background. We model an unabsorbed component at 0.16 keV corresponding to the Local Hot Bubble, and an absorbed component at 0.35 keV corresponding to the galactic halo emission. This model provides a good fit for all 4 quadrants, and we use the variation in the best fit parameters in each of the 4 quadrants as an estimate of the systematic uncertainty on the measurement of the soft background components when performing our spectral fits. 

For each spectral extraction region, our complete background model is therefore an absorbed powerlaw for the unresolved CXB, to which we add the contribution in each region from the resolved point sources that are seen in the XMM observations, added to the soft components measured from the RASS data.

\subsubsection{\textit{Suzaku} spectral fitting}
\label{Suzaku_spec_fit}
As shown in Fig. \ref{fig: A2199_mosaics}, we divide the cluster into 10 sectors and extract spectra in these sectors. For each region, we fit the XIS0, XIS1 and XIS3 spectra simultaneously in \textsc{xspec} using the extended C-statistic. For each spectrum, the non X-ray background spectrum is obtained using \textsc{xisnxbgen} and is subtracted. We then model the background using the background model described in Section \ref{Suzaku_bkg_mod}. We fit for the cluster emission with an absorbed \textsc{apec} model, with a hydrogen column of $0.08\times10^{22}$ cm$^{-2}$ (\citealt{LABsurvey}). Following \citet{urban2017uniform}, we use the abundance tables of \citet{Asplund2009}.

Deprojected temperatures and densities were obtained by following a method similar to the XSPEC model \textsc{projct} and described in \citet{Walker2013_Centaurus}, in which each annulus is modelled as the superposition of the ICM from that annulus and those outside it, weighting by appropriate volume scaling factors. 

Any gaps in the mosaics are linearly interpolated over, and are taken into account when performing spectral analysis. The coverage of the \textit{Suzaku} mosaic is very high, and over 80 percent of the cluster within $r_{200}$ is covered. To fully appreciate how high this coverage is, the results need to be considered in the context of the other \textit{Suzaku} studies of nearby galaxy clusters, which were limited to narrow strips from the core to the outskirts. In the Perseus cluster (\citealt{Urban2014}) the coverage is only 30 percent of the cluster within $r_{200}$, in the Coma cluster (\citealt{simionescu2013thermodynamics}) it is 15 percent, in the Virgo cluster (\citealt{Simionescu17}) it is only around 5 percent, and in the Centaurus cluster (\citealt{Walker2013_Centaurus}) it is 10 percent.

\begin{figure}
	\includegraphics[width=\columnwidth]{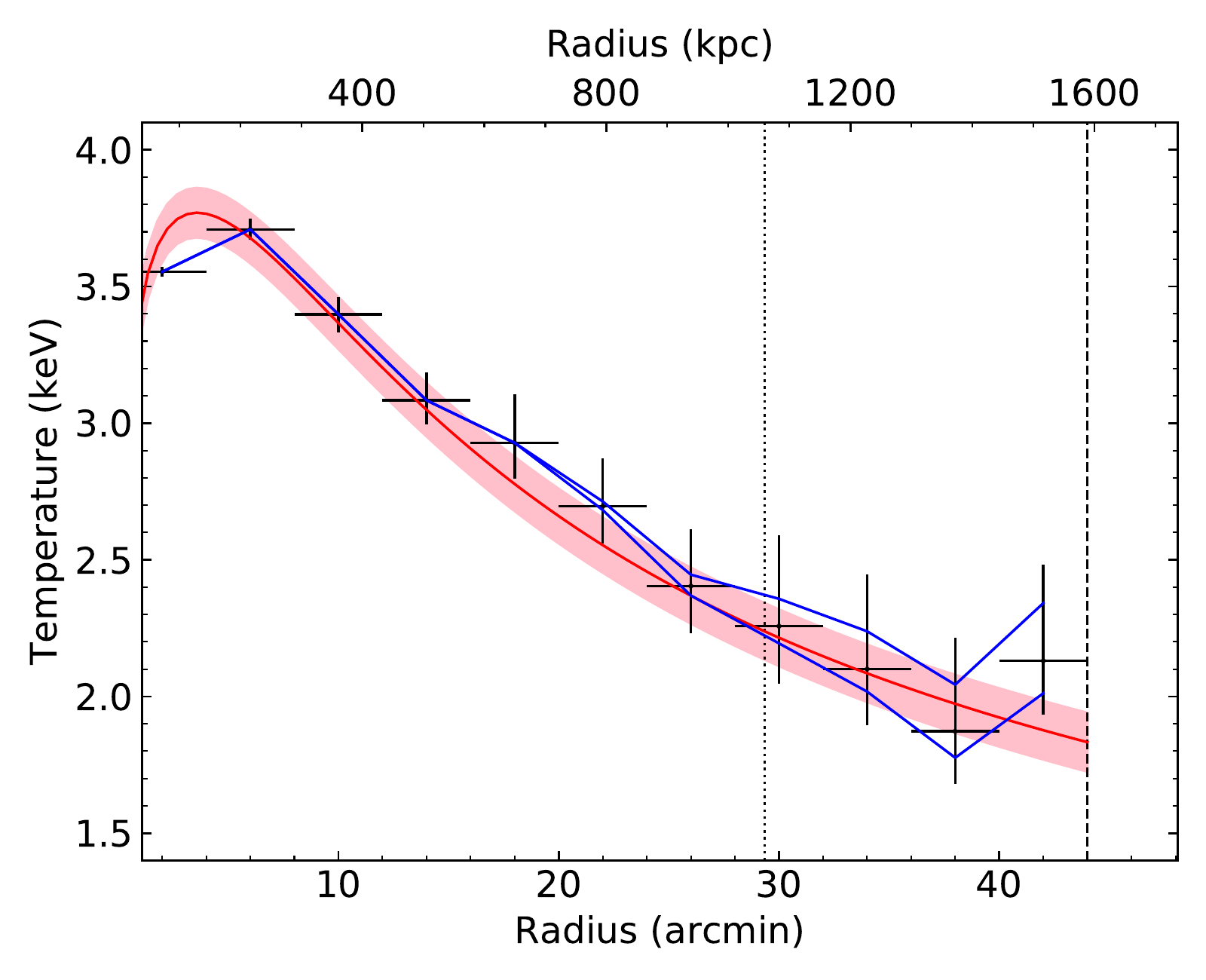}
	\caption{Azimuthally averaged temperature of the cluster A2199 recovered from the \textit{Suzaku} data. The solid blue line is the $1\sigma$ systematic error in the temperature measurements. For comparison, the red solid line is the parameterized temperature profile found by \citet{Ghirardini2019} by fitting the temperature measurements of a sample of 12 higher redshift clusters (the X-COP clusters) with the functional form of the temperature introduced by \citet{Vikhlinin2006}. The shaded pink area marks $1 \sigma$ error.} The vertical dotted and dashed lines mark, respectively, the locations of the $r_{500}$ and $r_{200}$ radii.
	\label{fig: A2199_temp_azav}
\end{figure}

\begin{figure}
	\includegraphics[width=\columnwidth]{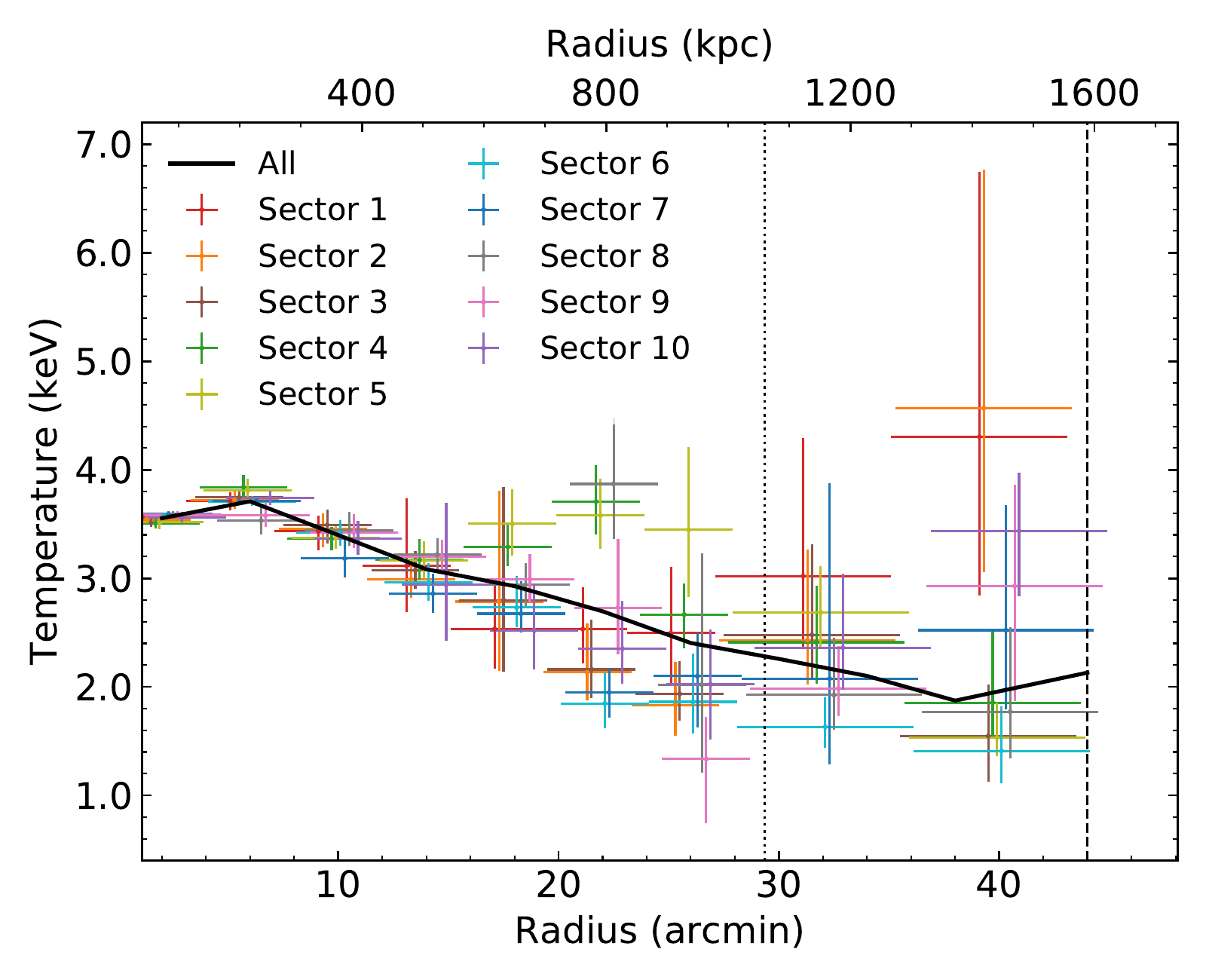}
	\caption{Same as Fig. \ref{fig: A2199_temp_azav}, except for azimuthal sectors. The thick black line shows the azimuthally averaged temperature profile. We introduce small offsets between data points at the same radius to aid readability.}
	\label{fig: A2199_temperature}
\end{figure}

The azimuthally averaged deprojected temperature profile is shown in Fig. \ref{fig: A2199_temp_azav}, while the deprojected temperature profiles in all 10 sectors are shown in Fig. \ref{fig: A2199_temperature}. All of the profiles we plot in this work have a linear radial axis to emphasize the outskirts. In Fig. \ref{fig: A2199_temp_azav}, we also show the parameterized temperature profile (the red solid line) obtained from \citet{Ghirardini2019} by fitting the temperature measurements of a sample of 12 higher redshift clusters (the X-COP clusters) with the functional form of the temperature introduced by \citet{Vikhlinin2006}. Our temperature measurements agree well with the best-fitting model of the temperature found by \citet{Ghirardini2019} over most of the radial range.

The metal abundance is also fit for, but we use larger azimuthal binning (we combine together adjacent sectors, thus dividing the cluster abundance measurements into 5 sectors rather than 10). When fitting for the metal abundance, we follow the methods described in \citet{urban2017uniform}. Namely, we only obtain metal measurements at radii greater than 0.25$r_{200}$ to avoid contamination from the central core. We also used as a threshold for measuring the outskirts metal abundance that the ICM signal to background ratio around the Fe-K line must be greater than 10 percent. We find that the outermost region we can study that satisfies this criterion is the 0.5-0.7$r_{200}$ annulus. The azimuthal variation of the metal abundance in this 0.5-0.7$r_{200}$ annulus is shown in Fig. \ref{fig: A2199_metal_outskirts}. We find that the metal abundance is relatively uniform with azimuth, and it has an average value of 0.29$^{+0.03}_{-0.03}$ $Z_{\odot}$. This is consistent with the results from studies of other \textit{Suzaku} clusters (\citealt{Werner2013}, \citealt{urban2017uniform}), which have found the metal abundance to be relatively uniform and consistent with a value of $0.3 \,Z_{\odot}$.

In each spectral fit, to determine the effect of systematic uncertainties we followed \citet{Walker2013_Centaurus} and produced 10000 realizations of the background model and the NXB level (which \citealt{Tawa2008} has found to have a systematic uncertainty of $\pm$3 percent), allowing all of these background components (unresolved CXB, Local Hot Bubble, Galactic Halo, and the NXB level) to vary simultaneously within their expected variances. We then performed the spectral fits for each realization of the background. The uncertainty in the background parameters is therefore folded through the deprojection, allowing a complete propagation of the systematic errors. Using the distribution of the 10000 spectral fits, we can that find the 1 $\sigma$ systematic errors, which we show as the solid blue lines in the temperature and entropy plots in Fig. \ref{fig: A2199_temp_azav} and Fig. \ref{fig: A2199_entropy_azav} respectively.

\section{Joint analysis}
\label{sec: joint_analysis}
In order to explore the thermodynamic properties of the cluster A2199, we combined our clumping corrected density measurements recovered from the \textit{XMM-Newton} observations with the deprojected gas temperature recovered from the \textit{Suzaku} observations. The entropy profile of A2199 is recovered through the gas temperature ($T$) and density ($n_{e}$) profiles as
\begin{equation}
    K = \frac{kT}{n_{e}^{2/3}}.
    \label{eq: entropy}
\end{equation}

The gas entropy is of particular interest since it tracks the thermal history of a cluster. In the presence of only non-radiative processes, numerical simulations \citep{Voit2005} predicted that the gas entropy outside the cluster core follows a power law with characteristic slope of 1.1,
\begin{equation}
 \frac{K}{K_{500}}=1.47\bigg(\frac{r}{r_{500}}\bigg)^{1.1},
 \label{eq: Voit_entropy}
\end{equation}
where the normalization is given by
\begin{equation}
 K_{500}=106\, {\rm{keV\, cm}}^2 \bigg(\frac{M_{500}}{10^{14}{\rm{M_\odot}}}\bigg)^{2/3}E(z)^{-2/3}f_b^{-2/3},
 \label{eq: K500}
\end{equation}
$M_{500}$ is the cluster mass at $r_{500}$, $f_b$ is the universal baryon fraction and assumed to be equal to 0.15, and $E(z)$ describes the redshift evolution of Hubble parameter. 

This baseline entropy profile tells us what the entropy of the ICM should be if only gravitational processes are considered. In addition to its shape, it also importantly gives a prediction for the normalization of the entropy. Any observed deviations from the predicted shape and normalization of the baseline entropy profile must be produced by non-gravitational processes, and so provide an insight into the history of heating of the ICM. 

On very small scales (much smaller than the scales we consider) in the very central cores of cool core clusters (the inner 10s of kpc), feedback from the central AGN causes the central entropy to be much flatter than the $r^{1.1}$ powerlaw, at around $r^{0.67}$ (e.g. \citealt{Babyk2018}). Outside the central 10s of kpc, the effect of the AGN on the entropy shape diminishes, and the entropy profile has been found to have a steeper gradient, typically close to but slightly less steep than an $r^{1.1}$ powerlaw (e.g \citealt{Babyk2018} find $K \propto R$ in the 0.1-1$r_{2500}$ region, corresponding to around the 0.03-0.3$r_{200}$ region). However, the normalization of observed entropy in the region  0.1-0.5$r_{200}$ has been found to be significantly higher than the baseline entropy profile prediction (e.g. \citealt{Pratt2010}, \citealt{Walker2019}), which is commonly called the entropy excess. This indicates that the central AGN may be able to increase the amount of entropy in the ICM above the baseline prediction out to around 0.5$r_{200}$, whilst the shape of the entropy profile retains a shape close to the expected $r^{1.1}$ powerlaw form.  

Around 0.6$r_{200}$, the entropy profiles of clusters tend to flatten slightly (e.g. \citealt{Pratt2010}, \citealt{Walker2019}), and the observed entropy level is seen to converge to the baseline entropy profile level. This indicates that the effect of the AGN on the entropy level in clusters can only reach as far out as 0.6$r_{200}$. The main focus of this paper is on the behaviour of the ICM in the outskirts, in the region between 0.6-1.0$r_{200}$. These regions are so far out that the central AGN cannot effect them, and any deviation of the observed entropy from the baseline entropy profile must be due to some other physical factor. 

The low X-ray surface brightness in these regions means that the gas temperature can only be measured through X-ray spectroscopy using an X-ray telescope with a low and stable X-ray background, and so accurate measurements of the temperature in these regions only became possible thanks to the \textit{Suzaku} observatory, whose low Earth orbit provided it with the requisite low and stable background level. Early observations of the outskirts of clusters with \textit{Suzaku} tended to find that the entropy in the 0.6-1.0$r_{200}$ region was below the expected baseline entropy profile, indicating that some additional physical process in the outskirts must be acting to reduce the observed entropy level (see \citealt{Walker2019} for a review). One important factor indicated by simulations is the effect of gas clumping; clumps of dense gas are expected to be more prevalent in the outskirts of clusters (\citealt{Nagai2011}). Such clumps are too faint to be resolved directly by \textit{Suzaku} (whose PSF is 2 arcmin half power diamater), and would act to bias the observations by raising the observed surface brightness and density, causing the density to be overestimated in the outskirts, and hence causing the entropy ($K = {kT}/{n_{e}^{2/3}}$) to be underestimated.    

Subsequent work using \textit{XMM-Newton} observations of the outskirts of clusters (which has a much better PSF than \textit{Suzaku} of around 15 arcsec half power diamater) has found that it is possible to measure the level of gas clumping in the outskirts by binning the XMM image into a Voronoi tesselation, and examining the distribution of the surface brightness within a given region of the cluster \citep{eckert2015gas,Tchernin2016,ghirardini2018xmm}. The gas clumping causes certain areas of the tessellation to have a higher than expected surface brightness, and this effect can be quantified by comparing the mean surface brightness in a region to the median surface brightness. The median is far less susceptible to outliers, and therefore provides a more realistic measure of the surface brightness, which is not affected by gas clumping. This allows the clumping factor to be found, as discussed in Section \ref{section:XMMreduction}, and allowing the densities we measure to be corrected for any clumping bias.

\begin{figure}
	\includegraphics[width=\columnwidth]{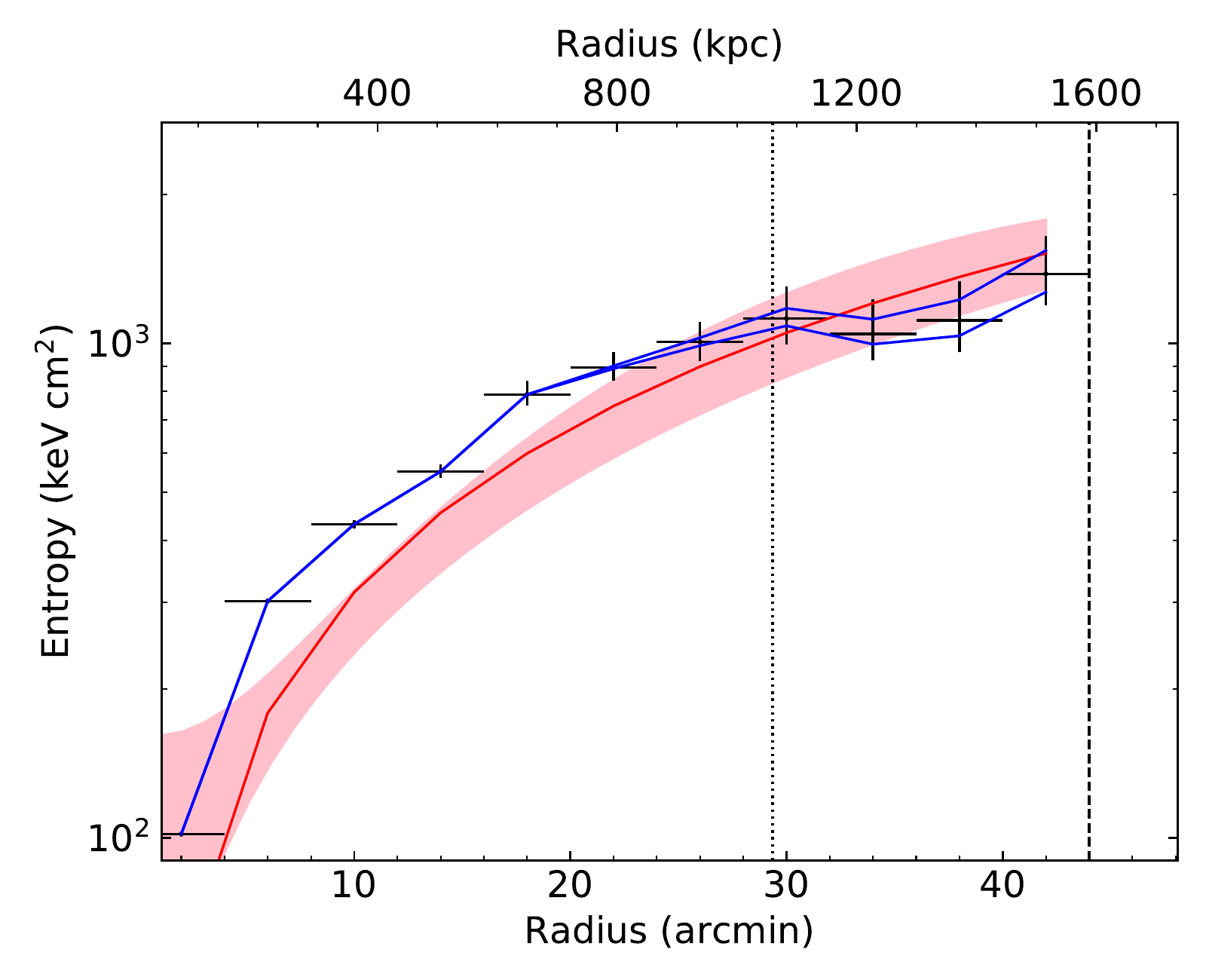}
	\caption{Clumping corrected entropy profile of the A2199 cluster obtained from the density and temperature measurements. The solid blue line is the $1\sigma$ systematic error in the entropy measurements. The red solid line marks the power-law entropy predicted by non-radiative simulations \citep{Voit2005}, and the shaded pink area represents the scatter from the non-radiative simulations. The vertical dotted and dashed lines mark the locations of the $r_{500}$ and $r_{200}$ radii, respectively.}
	\label{fig: A2199_entropy_azav}
\end{figure}

Fig. \ref{fig: A2199_entropy_azav} shows our measurements of the azimuthally averaged entropy profile of the A2199 cluster, which has been corrected for the effect of gas clumping. In the same figure, we also show the power-law entropy profile predicted by non-radiative simulations whose range is shown by the shaded pink region. Within $r_{500}$, our clumping corrected entropy measurements are in excess of the baseline entropy profile predicted by \citet{Voit2005}, as typically found for low mass clusters (\citealt{Pratt2010}). 

Beyond $r_{500}$, however, our clumping corrected measurements agree well with the baseline entropy profile predicted by non-radiative simulations (the pink shaded region in Fig. \ref{fig: A2199_entropy_azav}). This shows that, when the whole cluster is considered and corrected for gas clumping, the entropy level agrees with the level expected from purely gravitational structure formation.

To quantify the slope of our azimuthally averaged entropy measurements, we fitted the entropy profile with piecewise power laws in several radial ranges, as described in \citet{Ghirardini2019}. At radii between 2$-$26 arcmin, we find that the best fit yields a slope of $0.81\pm 0.19$. Between 26$-$34 arcmin, the slope computed from the piecewise power-law fits is $0.14 \pm 0.09$. Beyond 34 arcmin, however, the estimated slope is $1.20 \pm 0.23$, which is consistent with a slope of 1.1 predicted by non-radiative simulations.

\begin{figure}
	\includegraphics[width=\columnwidth]{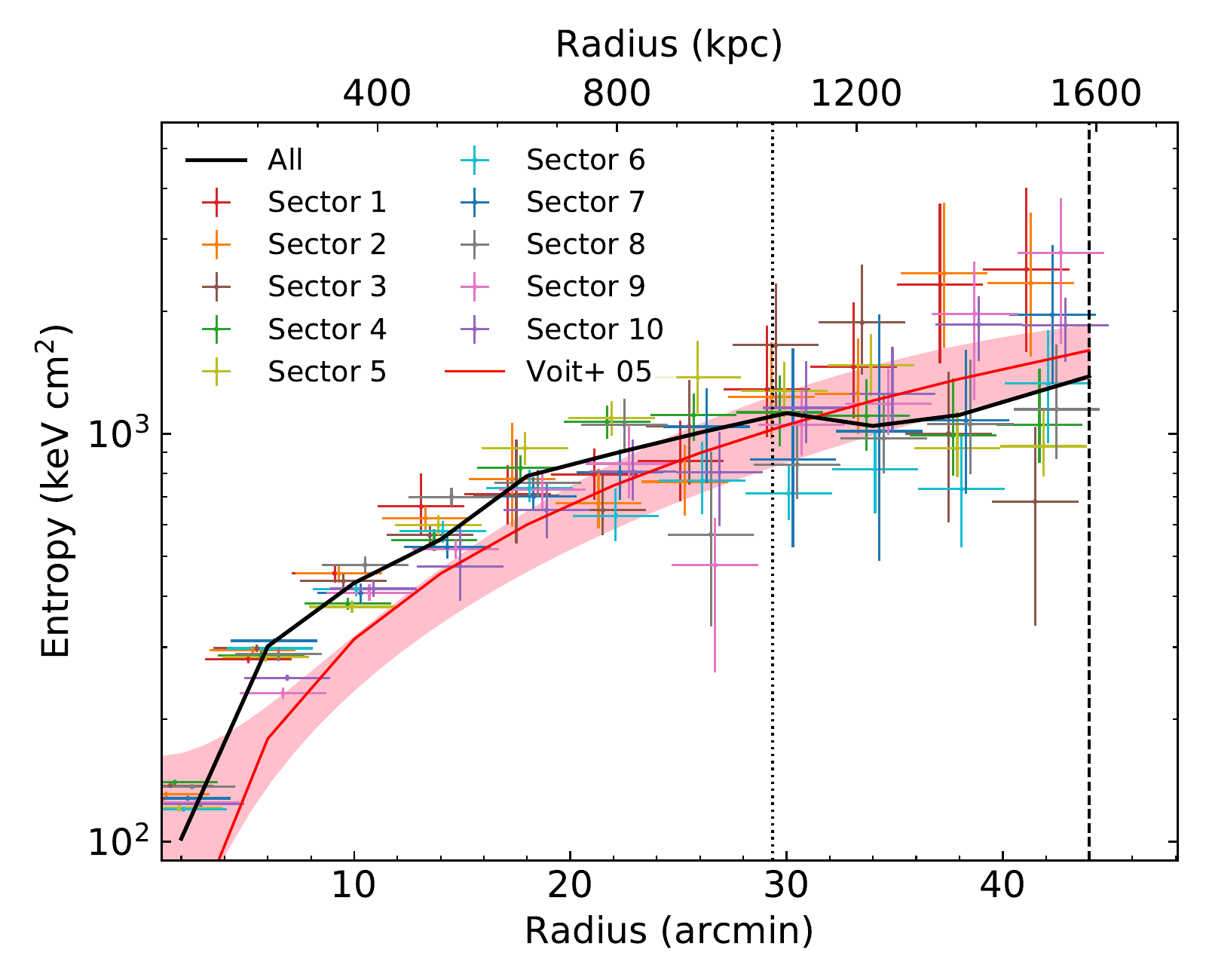}
	\caption{Same as Fig. \ref{fig: A2199_entropy_azav}, except for azimuthal sectors. The thick black line shows the azimuthally averaged entropy profile. To aid readability, small offsets are introduced between data points at the same radius.}
	\label{fig: A2199_entropy}
\end{figure}

In Fig. \ref{fig: A2199_entropy}, we present our clumping corrected entropy profiles in angular sectors. Beyond $r_{500}$, the entropy measurements in sectors show a significant azimuthal variation, with some of them lying significantly below the baseline entropy profile. These deviations from the baseline entropy profile suggest that, for a given sector considered by itself, some other physical process is affecting the entropy level. In particular, we see that the entropy in the outermost annulus of sector 3 lies well below the baseline entropy profile range. One candidate for causing such a low entropy in just one sector is the effect of filamentary gas streams, which are streams of low entropy gas seen in simulations \citep{zinger2016role}, flowing into the cluster along the direction of the cosmic web filaments which connect the cluster to the cosmic web.

We also calculated the gas and total masses of A2199 from the recovered density and temperature measurements using non-parametric methods, i.e., no modelling is adopted for the ICM properties. For the gas mass, the density measurements is integrated over a given volume and multiplied by the mean mass per electron, $\mu_e m_{\rm{p}}$. The gas mass takes the form:
\begin{equation}
 M_{\rm{gas}}=4\pi \mu_{\rm{e}} m_{\rm{p}} \int n_{\rm{e}}(r)r^2dr,
 \label{eq: gas_mass}
\end{equation}
where the integration is computed using a four-point spline interpolation and the limit is taken out to some cut-off radius.

The total mass, which is mainly made up of the dark matter, is calculated assuming that the ICM gas is in the hydrostatic equilibrium in the cluster's gravitational potential. The total mass can be expressed as
\begin{equation}
 M_{\rm{tot}}=-\frac{r^2}{G \mu_{\rm{e}} m_{\rm{p}} n_{\rm{e}}(r)} \frac{{\rm{d}}P_{\rm{g}}(r)}{{\rm{d}}r},
 \label{eq: tot_mass}
\end{equation}
where $G$ is the Newtonian gravitational constant, and the gas pressure, $P_{\rm{g}}$ ($= 1.83 n_{e} kT$), satisfies the ideal gas law. The derivative in equation (\ref{eq: tot_mass}) is computed using a three-point quadratic interpolation.

\begin{figure}
	\includegraphics[width=\columnwidth]{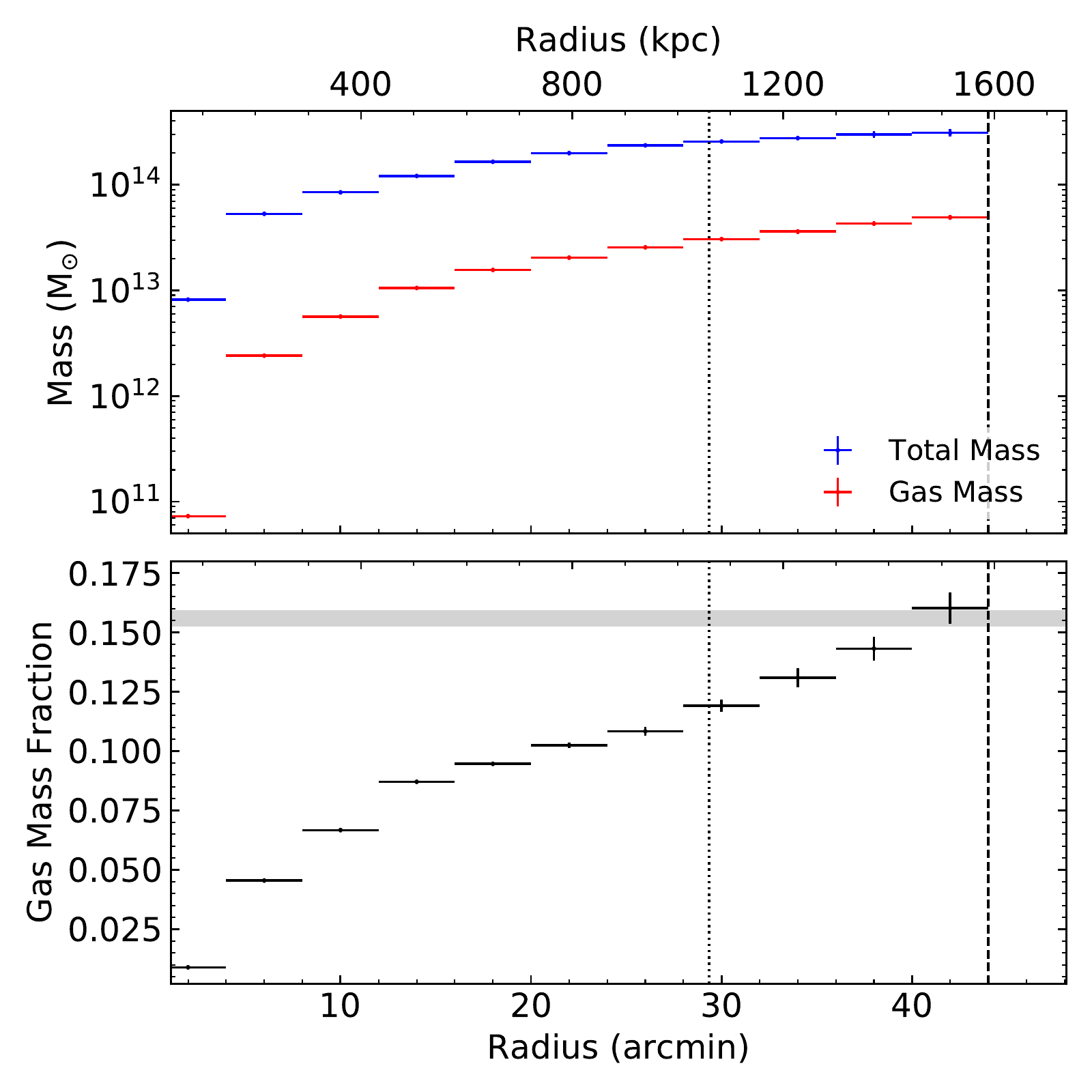}
	\caption{\textit{Top}: Radial profiles of the azimuthally averaged gas and total masses of A2199. \textit{Bottom}: Azimuthally averaged gas mass fraction profile of A2199. The gray area shows the universal baryon fraction \citep{ade2016planck}. The vertical dotted and dashed lines represent the locations of the $r_{500}$ and $r_{200}$ radii, respectively.}
	\label{fig: A2199_masses_azav}
\end{figure}

\begin{figure}
	\includegraphics[width=\columnwidth]{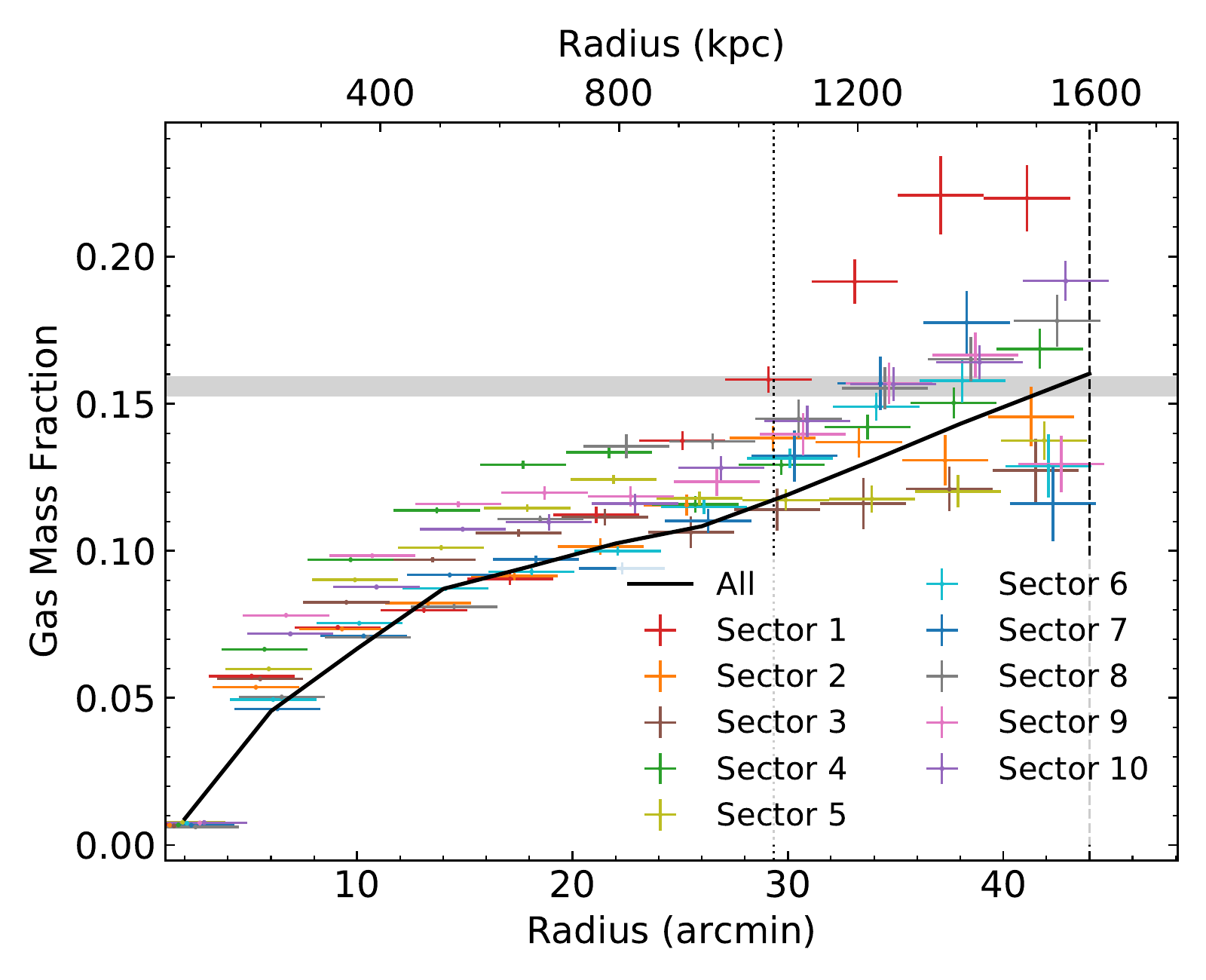}
	\caption{Gas mass fraction measurements in all 10 angular sectors of the A2199 cluster. The gray area shows the universal baryon fraction \citep{ade2016planck}. The vertical dotted and dashed lines mark the locations of the $r_{500}$ and $r_{200}$ radii, respectively.}
	\label{fig: A2199_gf}
\end{figure}

From these mass measurements, the gas mass fraction is then recovered as $f_{\rm{gas}}=M_{\rm{gas}}/M_{\rm{tot}}$. In Fig. \ref{fig: A2199_masses_azav}, we show the gas and total masses, and the gas mass fraction of A2199 out to the virial radius. We find that our total mass measurement within $r_{500}$ is $(2.41 \pm 0.15) \times 10^{14} \,\rm{M_\odot}$, consistent with the total mass of $(2.08 \pm 0.28) \times 10^{14} \,\rm{M_\odot}$ derived from the $M-T$ scaling relation \citep{lovisari2015scaling}. Within $r_{200}$ ($\approx 1.60$ Mpc), the derived value of the total mass is $(3.10 \pm 0.25) \times 10^{14} \,\rm{M_\odot}$. This value consistent with the total mass of ($3.16 \pm 0.48) \times 10^{14} \,\rm{M_\odot}$ measured within 1.65 Mpc using the \textit{Wide-field Infrared Survey Explorer} data \citep{lee2015galaxy}. Our computed gas mass fraction at $r_{200}$ is $0.160 \pm 0.008$, and agrees well with the universal baryon fraction of $0.156 \pm 0.003$ derived by \citet{ade2016planck}. The gas mass fraction profiles in azimuthal sectors are shown in Fig. \ref{fig: A2199_gf}. We note that the measurements in sectors show a significant deviation from azimuthally averaged gas mass fraction over most of the cluster's radial range. 

\section{Discussion}
\label{sec: discussion}

\subsection{Azimuthal variation in sectors}
\begin{figure}
	\includegraphics[width=\columnwidth]{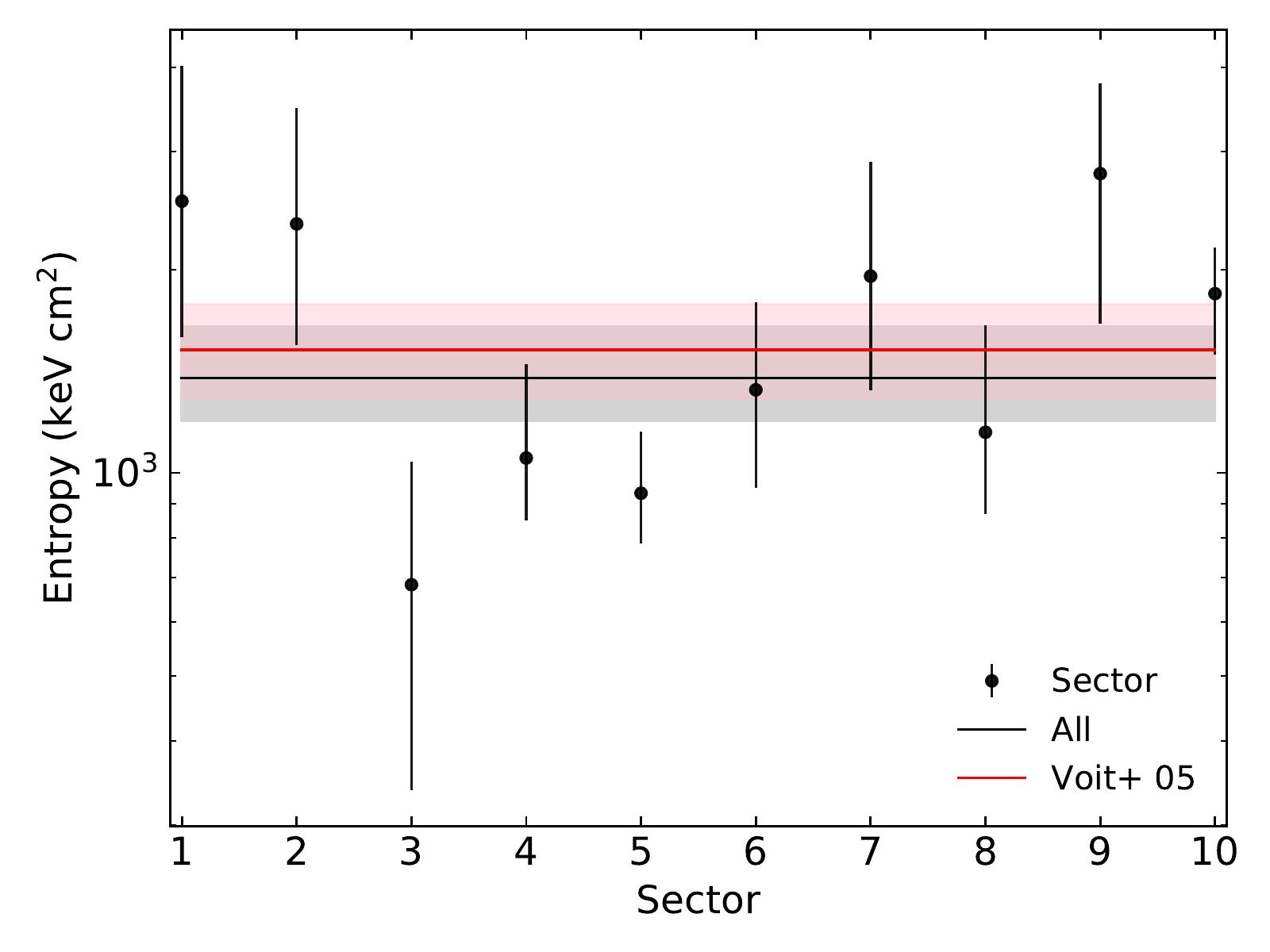}
	\caption{Azimuthal variation in the entropy measurements in the last annulus for all of the angular sectors. The black solid line is the azimuthally averaged entropy in the last bin, and the gray area marks $1\sigma$ uncertainty. The red solid line and the pink area represent, respectively, the power-law entropy and scatter in the last bin predicted by non-radiative simulations in \citet{Voit2005}.}
	\label{fig: A2199_entropy_last_bin}
\end{figure}

\begin{figure}
	\includegraphics[width=\columnwidth]{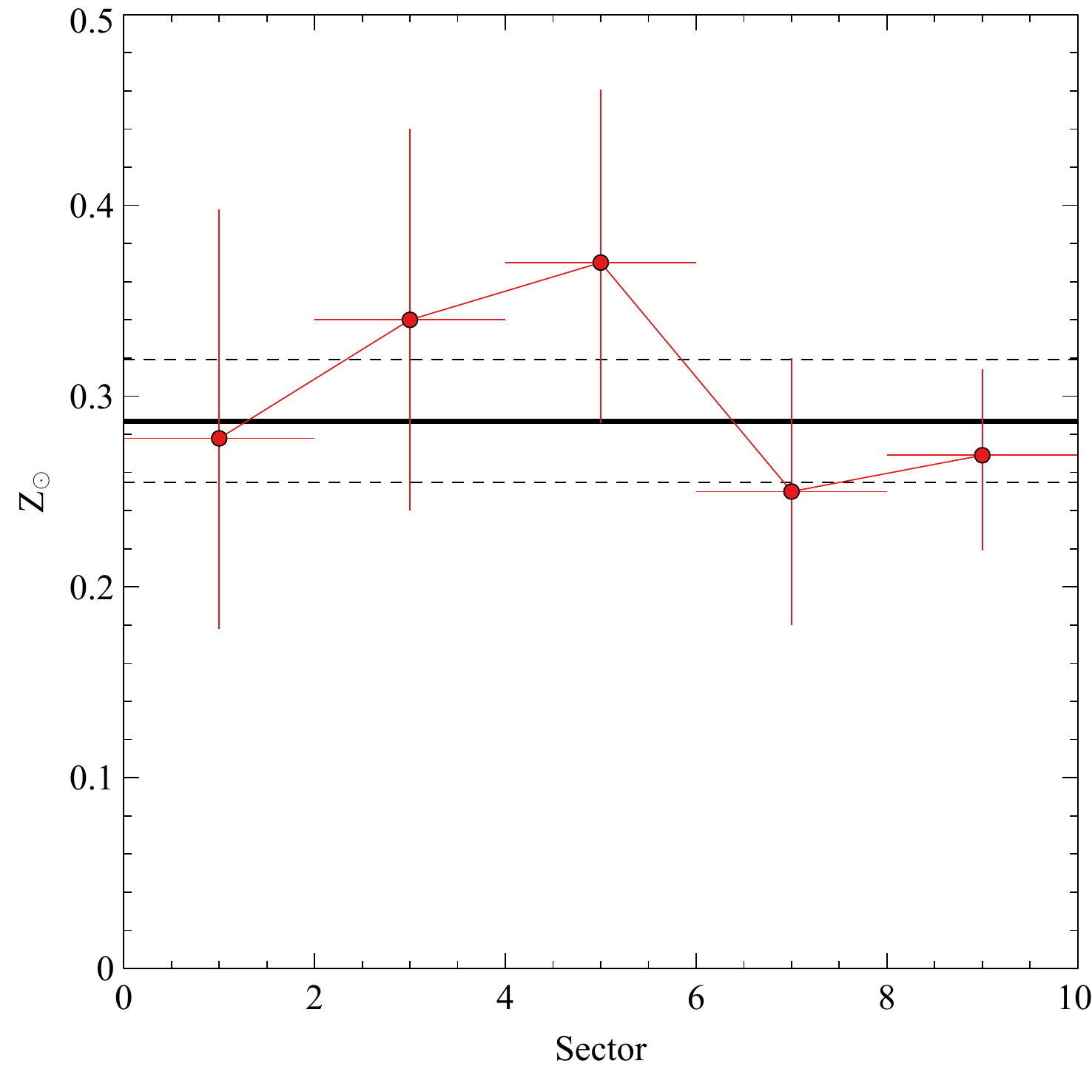}
	\caption{Metal abundance in the radial range 0.5-0.7$r_{200}$ for all of the sectors. The metal abundance appears relatively uniform in the outskirts, and the azimuthally averaged metal abundance  (shown by the solid horizontal black line, with the 1-$\sigma$ range given by the horizontal dashed black lines) is consistent with the value of 0.3 Z$_{\odot}$.}
	\label{fig: A2199_metal_outskirts}
\end{figure}

Fig. \ref{fig: A2199_entropy_last_bin} illustrates the azimuthal variation of the entropy in the last bin for all of the sectors. Sectors 3, 5, and 9 show a significant deviation from the azimuthally averaged entropy and with the theoretical expectations from purely gravitational collapse. Fig. \ref{fig: A2199_metal_outskirts} shows the metal abundance in the radial range 0.5-0.7$r_{200}$. There is no sign of significant deviation of the metal abundance from the azimuthally averaged profile in the outskirts.

We also investigated the level of asymmetry of the thermodynamic properties of the ICM in the entire cluster's radial range by computing the azimuthal scatter \citep{vazza2011scatter,ghirardini2018xmm}. The azimuthal scatter at radius $r$ is defined as
\begin{equation}
 \sigma_q(r)=\sqrt{ \frac{1}{N}\sum\limits_{i=1}^N\bigg(\frac{q_i(r)-\bar{q}(r)}{\bar{q}(r)}\bigg)^2},
 \label{equ: azimuthal_scatter}
\end{equation}
where $q_i(r)=\{C, n_e, T, K, M_{\rm{gas}}, M_{\rm{tot}}, f_{\rm{gas}}\}$ in sector $i$, $N$ is the number of sectors, and $\bar{q}(r)$ is the azimuthally averaged profile taken over the entire cluster volume.

\begin{figure}
	\includegraphics[width=\columnwidth]{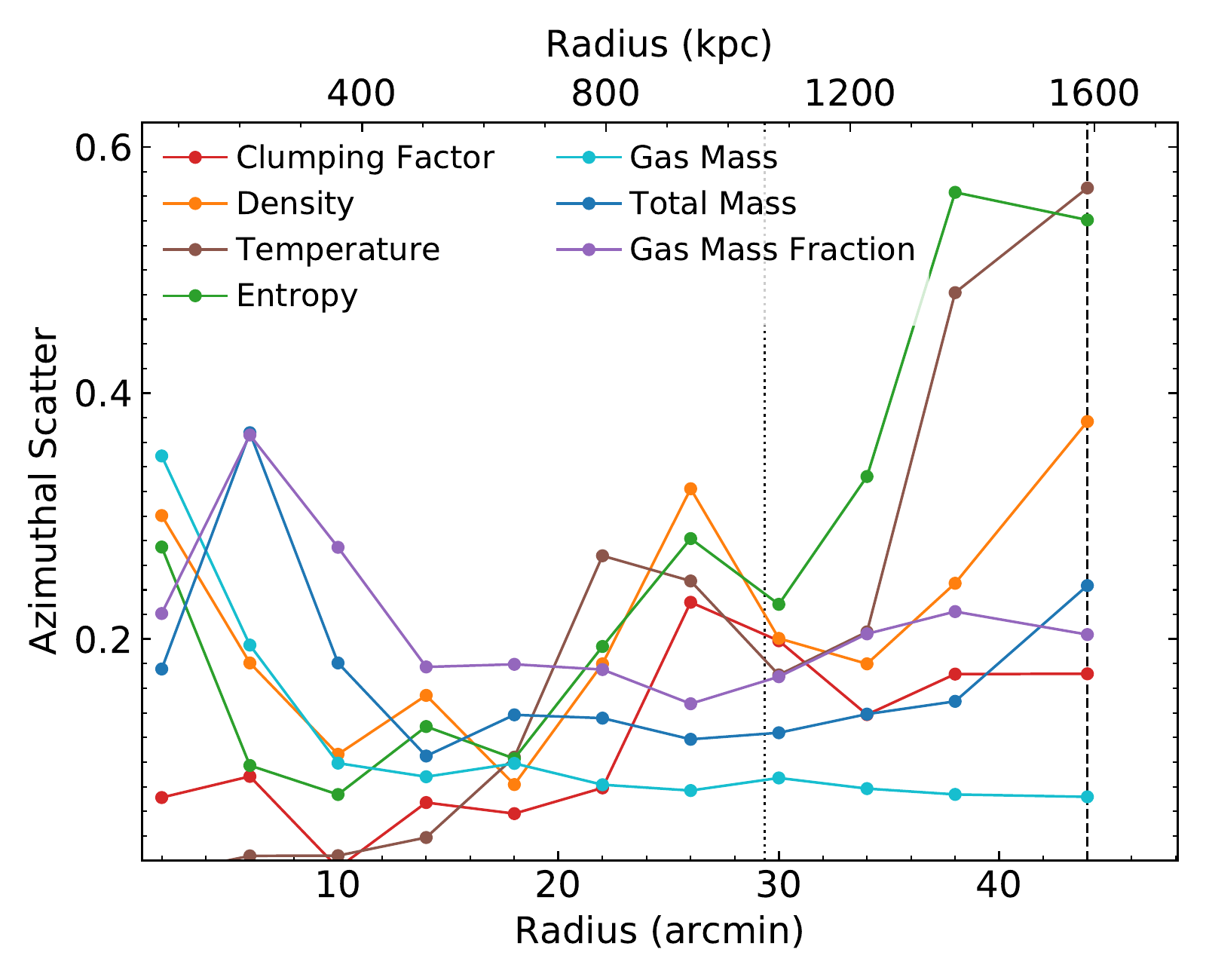}
	\caption{Azimuthal scatter profiles in the recovered thermodynamic properties of the A2199 cluster. A large value of the azimuthal scatter suggest the presence of an asymmetric gas. The vertical dotted and dashed lines mark the locations of the $r_{500}$ and $r_{200}$ radii, respectively.}
	\label{fig: A2199_az_scatter}
\end{figure}

In Fig. \ref{fig: A2199_az_scatter}, we plot the azimuthal scatter in the radial profiles of the recovered thermodynamic quantities of the A2199 cluster. A large value of $\sigma_q(r)$ seen in the cluster's core and outskirts for some thermodynamic quantities implies the presence of an asymmetric gas. Also, there is a location near a radial range of 20-30 arcmin where we see a jump in the azimuthal scatter of the gas density, temperature, and entropy of the ICM.  

Interestingly, the azimuthal scatter in the temperature is relatively high at r$_{200}$, where we find a value of 0.5, which is higher than the values found in the simulations of \citet{vazza2011scatter}, which find an azimuthal scatter in the temperature in the range 0-0.2. 

\subsection{Thermodynamics in outskirts}
\subsubsection{Entropy}
In the north and northeast directions (sectors 3-5), there is an apparent entropy deficit near the virial radius relative to the power-law entropy predicted from purely gravitational collapse, as shown in Figs. \ref{fig: A2199_entropy} and \ref{fig: A2199_entropy_last_bin}. In those directions, A2199 is connected kinematically to the cluster A2197 \citep{rines2001x} through a filamentary gas. These authors also found that two X-ray groups, positioning outside the virial radius of A2199, are connected to the A2199 cluster from the south direction. However, we find no sign of entropy decrement in this direction. 

Such low-entropy gas in the north and northeast directions is consistent with what is expected for a filamentary gas stream \citep{zinger2016role} with low entropy penetrating deep into the inner region of a cluster. However, we did not find a clear entropy deficit along those directions below $r_{500}$. \citet{zinger2016role} argued that the penetration of a filamentary gas stream into the inner region of a cluster may change over time, depending on a range of factors such as the mass accretion rate along the filament. It is possible that the mass inflow rate is low along those directions and, therefore, unable to penetrate deeper into the inner two-third of the cluster's virial radius.  

These results highlight the importance of having high azimuthal coverage of clusters in the outskirts. The results show that individual small sectors are not necessarily representative of the cluster as a whole, and that a spatially resolved study of the thermodynamics of the cluster in multiple sectors is needed to capture the full picture of what is happening in the outskirts.

\subsubsection{Metallicity}
\citet{Werner2013} found that the iron abundance in the outskirts of the nearby Perseus cluster is remarkably constant, as a function of radius and azimuth, statistically consistent with a constant value of $Z=0.306 \pm 0.012\, Z_{\odot}$ out to the virial radius of the cluster. More recently, \citet{urban2017uniform} reported metallicity measurements outside the central regions of a sample of 10 massive, nearby galaxy clusters observed with \textit{Suzaku}. Their results also revealed a uniform iron abundance in the outskirts, confirming the findings for the Perseus cluster. 

In the case of the A2199 cluster, the metal abundance for all of the azimuthal sectors appears relatively uniform in the outskirts (see Fig. \ref{fig: A2199_metal_outskirts}). \citet{Werner2013} argued that the homogeneous distribution of the metal-enriched gas in the outskirts requires that most of the intragalactic medium enriched by metal before galaxy clusters formed, likely more than 10 billion years ago, during the epoch of utmost star formation and black hole activity. Recent numerical simulations by \citet{biffi2017history} supported the idea that the wide spreading of the metal-enriched gas in the outskirts occurred mostly at early times ($z>2$), and found that this wide and homogeneous spreading of the metal-enriched gas is mainly driven by early AGN feedback acting on small haloes at $z>2$ with shallower potential wells.

\section{Conclusions}
\label{sec: conclusion}
We have presented a joint \textit{Suzaku} and \textit{XMM-Newton} analysis of the outskirts of the rich and nearby galaxy cluster A2199, the only nearby cluster to be observed with near complete azimuthal coverage with \textit{Suzaku}. By combining the clumping corrected density measurements obtained from the deprojection of the \textit{XMM-Newton}'s surface brightness with the deprojected temperature recovered from \textit{Suzaku}, we have been able to recover the entropy and the gas mass fraction measurements out to the virial radius. We have also measured the spatial variation of the metal abundance, a powerful probe of feedback physics, of the ICM in the A2199 outskirts.     

We find that the azimuthally averaged, clumping corrected entropy profile is consistent in the outskirts with the baseline entropy profile, and the average gas mass fraction in the outskirts is consistent with the mean cosmic baryon fraction. When split into 10 sectors however, the entropy and temperature demonstrate significant azimuthal variation. We perform an azimuthally resolved study of the outskirts metal abundance and find it to be consistent with being uniform with an average value of $0.29^{+0.03}_{-0.03}\, Z_{\odot}$, which agrees well with previous findings by \citet{Werner2013} and \citet{urban2017uniform}, and supports the hypothesis that the gas accreting onto galaxy clusters has been pre-enriched with metals.

\section*{Acknowledgements}
We thank the referee for helpful comments that improved the paper. Based on observations obtained with \textit{Suzaku}, a joint JAXA and NASA mission, and \textit{XMM-Newton}, an ESA science mission with instruments and contributions directly funded by ESA Member States and NASA.  

\section*{Data Availability}

The \textit{Suzaku} data used in the paper are publicly available from the High Energy Astrophysics Science Archive
Research Centre (HEASARC) Archive.
The \textit{XMM-Newton} Science Archive (XSA) stores the archival data used in this paper, from which the data are publicly available for download.  The XMM data were processed using the \textit{XMM-Newton} Science Analysis System (SAS). The
software packages heasoft and xspec were used, and these can be downloaded from the HEASARC software web-page. Analysis and figures were produced using \textsc{python} version 3.7.

\bibliographystyle{mn2e}
\bibliography{A2199}

\appendix

\section{\textit{Suzaku} and \textit{XMM-Newton} Observations of A2199}

\begin{table*}
    \centering
    \caption{\textit{Suzaku} observations of A2199}
    \begin{tabular}{lcccccc}
    \hline
    Name        & RA     & Dec.  & Obs. Date   & Obs. ID     & Exposure (ks)    & Offset (arcmin)      \\
        \hline
A2199 EAST OFFSET   & 16 27 03.98 & +40 14 14.6 & 2010-09-19  & 805042010 & 58.44 & 46.350 \\
A2199 NE            & 16 30 03.84 & +39 47 02.8 & 2014-01-04  & 808050010 & 44.16 & 22.849 \\
A2199 FE            & 16 33 08.52 & +39 38 17.9 & 2014-01-05  & 808051010 & 40.35 & 52.760 \\
A2199 OFFSET8       & 16 31 08.90 & +40 00 38.9 & 2011-08-23 & 806154010 & 30.71 & 41.275 \\
A2199 OFFSET6       & 16 29 06.05 & +38 47 10.0 & 2011-08-19 & 806152010 & 29.79 & 44.657 \\
A2199 OFFSET7       & 16 31 29.14 & +39 03 44.3 & 2011-08-19 & 806153010 & 28.98 & 43.336 \\
A2199 OFFSET13      & 16 29 36.19 & +39 59 42.4 & 2011-09-11 & 806159010 & 28.65 & 30.444 \\
A2199 OFFSET16      & 16 31 11.86 & +39 41 45.6 & 2011-10-06 & 806162010 & 28.32 & 31.551 \\
A2199 OFFSET15      & 16 25 55.68 & +39 53 20.4 & 2011-09-19 & 806161010 & 28.17 & 37.962 \\
A2199 OFFSET9       & 16 26 28.56 & +38 54 49.7 & 2011-09-12 & 806155010 & 28.13 & 44.286 \\
A2199 OFFSET14      & 16 25 02.50 & +39 31 16.3 & 2011-09-21 & 806160010 & 27.44 & 41.366 \\
ABELL 2199 OFFSET3 & 16 29 26.30 & +39 18 59.4 & 2006-10-04  & 801059010 & 25.25 & 15.695 \\
ABELL 2199 CENTER   & 16 28 46.13 & +39 29 02.4 & 2006-10-01  & 801056010 & 24.93 &  2.998 \\
ABELL 2199 OFFSET1 & 16 28 06.46 & +39 39 01.4 & 2006-10-01  & 801057010 & 24.63 &  9.578 \\
ABELL 2199 OFFSET4 & 16 28 07.10 & +39 18 58.0 & 2006-10-04  & 801060010 & 24.45 & 13.770 \\
A2199 OFFSET12      & 16 27 41.57 & +39 55 45.5 & 2011-09-16 & 806158010 & 23.59 & 26.527 \\
ABELL 2199 OFFSET2 & 16 29 26.86 & +39 38 58.2 & 2006-10-03  & 801058010 & 22.71 & 12.189 \\
A2199 OFFSET3       & 16 26 29.81 & +39 33 53.3 & 2011-08-17 & 806149010 & 21.37 & 24.640 \\
A2199 OFFSET11      & 16 26 27.26 & +39 13 17.8 & 2011-09-11 & 806157010 & 20.65 & 30.964 \\
A2199 OFFSET1       & 16 29 52.68 & +38 59 59.3 & 2011-08-16 & 806147010 & 20.30 & 34.720 \\
A2199 OFFSET2       & 16 28 05.33 & +38 59 40.9 & 2011-08-17 & 806148010 & 20.29 & 32.371 \\
A2199 OFFSET4       & 16 31 03.89 & +39 22 11.3 & 2011-08-18 & 806150010 & 18.71 & 29.836 \\

        \hline
    \end{tabular}
    \label{tab: Suzaku_observations}
\end{table*}

\begin{table*}
    \centering
    \caption{\textit{XMM-Newton} observations of A2199}
    \begin{tabular}{lcccccc}
    \hline
    Name   & RA     & Dec.    & Obs. Date  & Obs. ID     & Exposure (ks)  & Offset (arcmin)      \\
        \hline
A2199 1   & 16 28 37.56 & +39 33 09.4 & 2002-07-04 & 0008030201  & 22.98 & 1.69 \\
A2199 2   & 16 30 32.61 & +39 23 03.0 & 2011-09-08 & 0671900101  & 31.02 & 23.85 \\
A2199 3   & 16 26 39.40 & +39 00 53.1 & 2013-02-22 & 0691010101  & 47.83 & 38.12 \\
A2199 4   & 16 25 43.42 & +39 26 24.7 & 2013-02-28 & 0691010901  & 31.91 & 33.87 \\
A2199 5   & 16 26 39.40 & +39 00 53.1 & 2013-02-26 & 0691011001  & 33.92 & 38.12 \\
A2199 6   & 16 28 38.19 & +39 33 01.7 & 2013-08-12 & 0723801101  & 57.00 & 1.58 \\
A2199 7   & 16 26 44.30 & +39 26 38.0 & 2017-03-03 & 0784521101  & 40.00 & 22.28 \\
A2199 8   & 16 30 32.00 & +39 44 52.0 & 2017-03-07 & 0784521201  & 40.00 & 25.88 \\
A2199 9   & 16 28 38.19 & +39 58 16.0 & 2017-03-18 & 0784521301  & 30.00 & 26.80 \\
A2199 10  & 16 30 44.50 & +40 06 05.1 & 2017-03-21 & 0784521401  & 37.80 & 42.33 \\
A2199 11  & 16 26 44.30 & +39 53 01.0 & 2017-03-21 & 0784521501  & 32.00 & 30.57 \\
A2199 12  & 16 24 57.99 & +39 06 07.0 & 2017-03-22 & 0784521701  & 31.90 & 49.37 \\

  \hline
    \end{tabular}
    \label{tab: xmm_observations}
\end{table*}

\end{document}